\documentclass[onecolumn,usenatbib]{mnras}
\usepackage{graphicx}
\usepackage[export]{adjustbox}
\usepackage{xcolor}
\newcommand{\Dpar}{D_{\parallel}}
\newcommand{\Dperp}{D_{\perp}}

\newcommand{\lskip}{\vskip \baselineskip}
\newcommand{\nskip}{\lskip \noindent}
\newcommand{\bm}[1]{\mbox{\boldmath$ #1 $}}
\newcommand{\grad}{\mbox {\boldmath $\nabla$}}
\newcommand{\bdot}{\mbox{$\bm{\: \cdot \:}$}}
\newcommand{\halfskip}{\vskip 0.5\baselineskip}
\newcommand{\be}{\halfskip \begin{equation}}
\newcommand{\ee}{\end{equation} \halfskip \noindent}
\newcommand{\ba}{\halfskip \begin{eqnarray}}
\newcommand{\ea}{\end{eqnarray} \halfskip \noindent}

\newcommand{\hatz}{\bm{\hat{z}}}
\newcommand{\hatr}{\bm{\hat{r}}}

\newcommand{\hatphi}{\bm{\hat{\phi}}}
\newcommand{\hatb}{\bm{\hat{b}}}
\newcommand{\nus}{\nu_{\rm s}}
\newcommand{\btimes}{\bm{\times}}

\begin{document}
\title[Propagation of Galactic cosmic rays: effect of drift and winds]
{The effects of drift and winds on the propagation of  Galactic cosmic rays}
\author[AL-Zetoun \& Achterberg]{A. AL-Zetoun$^1$\thanks{E-mail: a.al-zetoun@astro.ru.nl}  \&     A. Achterberg$^1$ \\  
$^1$Department of Astrophysics, IMAPP, Radboud University, Nijmegen, P.O. Box 9010, 6500 GL Nijmegen, The Netherlands}

\date{Accepted..., Received...; in original form ...}
\pubyear{2018}
%\color{red}
\maketitle
\begin{abstract}
{ 

We study the effects of drift motions and the advection by a Galactic wind on the propagation of cosmic rays in the Galaxy. 
We employ a simplified magnetic field model, based on (and similar to) the Jansson-Farrar model for the Galactic magnetic field.
Diffusion is allowed to be anisotropic. The relevant equations are solved numerically, 
using a set of stochastic differential equations. Inclusion of drift and a Galactic wind significantly shortens the residence time of
cosmic rays, even for moderate wind speeds. 
}
\end{abstract}
%%%%%%%%%%%%%%%%%%%%%%%%%%%%%%%%%%%%%%%%%%%%%%

%%%%%%%%%%%%%%%%%%%%%%%%%%%%%%%%%%%%%%%%%%%%%%%

\section{Introduction}
\label{intro:1}

Cosmic rays (CRs) propagate in the Galaxy and through the surrounding  halo around the Galactic disk by a combination of diffusion, 
drift through the ambient magnetic field and advection by a large-scale wind e.g. \citet{StrongM2007}.
These processes are usually studied using by solving a diffusion-advection equation.
In addition CRs can gain (through re-acceleration) or lose (through expansion losses in a wind) energy during propagation.
During propagation CR composition is changed due to spallation on ISM nuclei or by radioactive decay of unstable nuclei. 
\nskip
The charged CR nuclei (and CR electrons and positrons) are collisionally coupled  
to a possible Galactic wind, causing them to be advected by the bulk flow, see for instance \citet{Skilling}.
This coupling is due to frequent scattering of the CRs as a result of wave-particle interactions with low-frequency MHD waves.
The intensity of these waves is determined by the CR density gradient, which
causes the excitation of Alfv\'en waves, see \citet{Wentzel} or \citet{Skilling}.
\nskip  
The  mechanisms  driving a  Galactic winds include the deposition of mechanical energy into the ISM by core collapse supernovae,
see for instance \citet{Martin99}, and the effects of radiation- or CR pressure e.g. \citet{Hopkins2012}. 
The effect of such a large-scale wind is now routinely included in numerical simulations of CR propagation. 
\nskip
In the diffuse Galactic disk 
there is a rough equipartition of the CR  energy density and the energy density of the Galactic magnetic field, see for example \citet{Beck2005}.
This implies that CRs can play a significant role in the dynamics of the ISM. That last  point will not be addressed in this paper.

\nskip
CR drift motions  with respect of the large-scale magnetic field
are usually a combination of gradient and curvature drifts. 
These are indispensable ingredients in the study the CR propagation in the Galaxy and (on a much smaller scale) in the Solar Wind. 
For example: \citet{Jokipii1977}  has presented a model of CR propagation in the solar wind that includes drift. 
Those authors  studied the effects of gradient drifts on CR transport, with 
the magnetic field taken to be an Archimedean spiral.

\nskip
In this paper we take into account the effects of cross-field drift in the curved magnetic field, and the effects of CR advection away from the disk  by the Galactic winds on the propagation of CRs in the Galaxy.

\nskip
A number of analytical models for the Galactic magnetic field have been published in recent years, see for instance: 
\citet{Sun2008},\citet{Jaffe2010}, and \citep{JF2012A,JF2012B}.
 \nskip
The \citep{JF2012A,JF2012B} model,  hereafter JF12, {\em does} include a detailed model for the vertical field. 
In this paper we use a simplified GMF model (see below), based on JF12 model. 
This model preserves most of the features of the JF12 model:
the field in the plane  of the disk (horizontal field) is essentially unchanged, but, close to the disk mid-plane, 
the vertical field is taken to be perpendicular to the Galactic disk. 
The reason for this approach is mainly that the simplified model, unlike the original
JF12 model,  allows a (relatively) simple analytical calculation of CR drifts, 
which can then be used to check the numerical results. 

\nskip
The reason for this approach is mainly that the simplified model, unlike the original JF12 model,
allows a (relatively) simple analytical calculation of CR drifts. These analytical results, summarized in
the Appendix, are used to calculate the drift speed in the advective step.
Other than the inclusion of CR drift and advection by a Galactic wind, the numerical methods used
in this Paper are identical to those used in the two previous papers ( \citet{ALZetoun2018}, and \citet{ALZetoun2020}). As a result, the performance
and efficiency of the code is comparable to what was found before.

\nskip
The rest of the paper is organized as follows:
In Section \ref{sec:2.1}, we describe  the large scale Galactic magnetic field model. We discuss  our propagation model  using relevant input, like the diffusion tensor, the path length  and the grammage distribution,  advection by Galactic wind, and drift velocity  in Section \ref{sec:2.2},  \ref{sec:2.3}, and \ref{sec:2.4}, respectively.
In Section \ref{sec:3} we discuss  the spatial distribution of CRs in the Galaxy when we include the drift motion and the advection by  Galactic wind. 
Finally,  Section \ref{sec:4} contains the conclusions. 
In the Appendix we give the details of the CR drift in the modified Jansson-Farrar field.

%%%%%%%%%%%%%%%%%%%%%%%%%%%%%%%%%%%%%%%%%%%%%%%%%%%%%%%%%%%%%%%%%%%%%%%%
%%%%%%%%%%%%%%%%%%%%%%%%%%%%%%%%%%%%%%%%%%%%%%%%%%%%%%%%%%%%%%%%%%%%%%%%%
\section{Simulation assumptions and  parameters}
\subsection{ the Galactic magnetic field model}
\label{sec:2.1}

We briefly discuss our modification of the GMF model of \citet{JF2012A} and \citet{JF2012B}.
The JF12 model has three distinct components: spiral disk field, a  poloidal X-shaped field, and  a toroidal halo field.
Our simplifications involve the disk component as well as the  {\em X-field} component, as explained immediately below.
\begin{enumerate}
\item For a distance $|z|  \le h = 0.4 \: {\rm kpc}$ from the disk mid-plane we take the field to be
\be
\label{Btotal}
\bm B(r) =  B_{0}^{\rm D} \; \left(\frac{{r_{0}}}{ r}\right) \;  \big(\sin p \;\hatr + \cos p \;\hatphi  \big) \;+\;   
B_{0}^{\rm X}\:   {\rm exp}(-r_{\rm p}/ H_{\rm X})  \;  \sin i(r_{\rm p}) \hatz \; .
\ee
Here the first term is the spiral field in the disk plane, while the second term is the vertical X-field.  
The spiral pitch angle $p = 11.5$ degrees and the value of $B_{0}^{\rm D}$ is different in the 8 spiral sections
of the field. The disk field scales with Galacto-centric radius r as: $\bm{B}^{\rm D}  \propto r^{-1}$, the radius $r_{0}$ can be chosen arbitrarily,   in our simulations  we use the value of   $r_{0} = 5 \: {\rm kpc}$, see \citet{JF2012A} and \citet{ALZetoun2018} for details. We take the $X$-field to be purely vertical with the same
properties as the vertical component of the JF12 field: $B_{0}^{\rm X} = 4.6 \: \mu {\rm G}$, $H_{X} = 2.9 \: {\rm kpc}$, and
$\tan i(r_{\rm p}) = (r_{\rm X}/r_{\rm p}) \: \tan i_{0}$ for $r_{\rm p} < r_{\rm X} = 4.8 \: {\rm kpc}$ and 
$\tan i(r_{\rm p}) = \tan i_{0}$ for $r_{\rm p} \geq r_{\rm X}$. Here $r_{\rm p} = r/(1 + h/r_{\rm X} \tan i_{0})$ for $r_{\rm p} < r_{\rm X}$
and $r_{\rm p} = r - h/\tan i_{0}$ for $r_{\rm p} \geq r_{\rm X}$  and $\tan i_{0} = 1.15$ ($i_{0} = 49$ degrees).

\item For $|z| > h$ we assume that the disk field vanishes abruptly. In the JF12 model this transition is more gradual.
The X-field remains in the form given in \citet{JF2012A}:
\be
	\bm{B}(r) = B^{\rm X}(r_{\rm p}) \: \cos(i) \: \hatr +  B^{\rm X}(r_{\rm p}) \: \sin(i) \:  \hatz \; .
\ee
We neglect the relatively weak halo field.
\end{enumerate}
%%%%%%%%%%%%%%%%%%%%%%%%%%%%%%%%%%%%%%%%%%%%%%%%%%%%%%%%%%%%%%%%%%%%%%%%%%%%
%%%%%%%%%%%%%%%%%%%%%%%%%%%%%%%%%%%%%%%%%%%%%%%%%%%%%%%%%%%%%%%%%%%%%%%%%%%%
\subsection{The equations for CR propagation}
\label{sec:2.2}
Recently, \citet{ALZetoun2018} presented the results from a fully three-dimensional simulation of CR propagation,  
based on  the It\^o formulation of the Fokker Planck in terms of a set of stochastic differential equations.  
The results allowed for anisotropic diffusion but neglected the effects of CR drift and the Galactic wind.
In this paper we include these effects.  
\nskip
In finite-difference form the  It\^o formulation advances the position $\bm{x}$ of a simulated CR as the sum of a regular advective
step and a diffusive stochastic (random) step. In a time span $\Delta t$ one has
\be
\label{TPmotion}
	\Delta \bm{x} = \left(\bm{V}_{\rm w}(\bm{x}) + \bm{V}_{\rm dr}(\bm{x}) \right)  \: \Delta t + \Delta \bm{x}_{\rm diff} \; .
\ee
The proper definitions of the wind speed $\bm{V}_{\rm w}$ and the drift speed $\bm{V}_{\rm dr}$ are given directly below.
The  diffusive step $\Delta \bm{x}_{\rm diff}$ has the form: 
\be
\label{diffstepsize}
	\Delta \bm{x}_{\rm diff} =    \sqrt{2 \Dperp \: \Delta t}\; \xi_{1}\: \hat{\bm{e}}_{1}
	+  \sqrt{2 \Dperp \: \Delta t}\: \xi_{2} \;	\hat{\bm{e}}_{2} + \sqrt{2 \Dpar \:  \Delta t}\; \xi_{3} \: \hat{\bm{e}}_{3} \; ,
\ee
It involves Gaussian random steps with rms size $\sqrt{2 \Dpar \: \Delta t}$ in the direction along the magnetic field,
and random steps with rms size $\sqrt{2 \Dperp \: \Delta t}$ in the two directions in the plane perpendicular to the magnetic field. For more details about these aspects of the model, see \citet{ALZetoun2018}.
To achieve this, the random variables $\xi_{1}$, $\xi_{2}$  and $\xi_{3}$ are independently drawn from a Gaussian distribution with zero mean and unit dispersion.  
In our simulations we use a constant value for the ratio $\Dperp/\Dpar \equiv \epsilon$. The scaling with CR rigidity ${\cal R} = pc/qB$
is $D_{\parallel} \propto {\cal R}^{\delta}$. Values quoted for $\Dpar$ are for protons with an energy of 1 GeV.
%%%%%%%%%%%%%%%%%%%%%%%%%%%%%%%%%%%%%%%%%%%%%%%%%%%%%%%%%%%%%%%%%%%%%%%%%%%%%%%%%%

%%%%%%%%%%%%%%%%%%%%%%%%%%%%%%%%%%%%%%%%%%%%%%%%%%%%%%%%%%%%%%%%%%%%%%%%%%%%%%
\subsection{Path length and the grammage distribution}
\label{sec:2.3}
The  path length distribution (PLDs)  is an important quantity that can be  determined from measurements of
the CR composition at Earth. It determines the number of spallation reactions that a typical primary CR undergoes, that can be
measured by using the ratio of fluxes of secondary-primary nuclei, like Boron to Carbon ratio.
In our calculation the path length increases by $\delta \ell = v \: \Delta t$ over a time span
$\Delta t$, with $v$ the instantaneous CR velocity. The grammage increases as:
\be
\label{grammage}
	\Delta \Sigma_{\rm cr} = \rho(\bm{r}_{\rm cr} ) \: v \: \Delta t \; , \: \mbox{with} \; 
	\rho(\bm{r}) = \left\{ \begin{array}{ll}
	{\displaystyle \rho_{0} \: {\rm exp}(-|z|/H_{\rm d}(r))} & \mbox{for $r < R_{\rm c}$}  \; , \\
	& \\
	{\displaystyle \rho_{0} \: {\rm exp}(-|z|/H_{\rm d}(r))} \: {\rm exp}\left[ \: -(r - R_{\rm c})/R_{\rm d} \: \right] & \mbox{for $r > R_{\rm c}$}  \; . \\
	\end{array} \right. 
\ee
\nskip
where $\rho(\bm{r})$ is the density of the diffuse gas at CR position $\bm{r}$,  
$v \simeq c $ is the velocity of the CR, and $\bm{r}_{\rm cr}$ is the instantaneous position of the CR inside the Galaxy. 
The radial scale length $R_{\rm c}$ in the density distribution 
equals $R_{\rm c} = 7$ kpc. The vertical density scale height $H_{\rm d}(r) = H_{0} \: {\rm exp}(r/R_{\rm h}) $. Here $R_{\rm h} \simeq 9.8$ kpc and $H_{0} \simeq 0.063$ kpc.

%%%%%%%%%%%%%%%%%%%%%%%%%%%%%%%%%%%%%%%%%%%%%%%%%%%%%%%%%%%%%%%%%%%%%%%%%%%%%%%
\subsection{Model for the Galactic wind}
\label{sec:2.4}

Several  theoretical papers e.g: \citet{Breitschwerdt1991}, \citet{Zirakashvili1996}, and \citet{Pakmor2016} 
conclude that CRs can play an important role in launching Galactic winds. For instance:
\citet{Breitschwerdt1991}, and \citet{Breitschwerdt1993} showed that the Galactic  winds are accelerated  
by the pressure of the CRs, as well as by gas- and MHD wave pressure.  As a result the wind velocity can reach several hundred $\rm \; km/s$. \citet{Everett2008} shows that the initial velocity, close to the disk, is about $200\; \rm km/s$ and increases to  $ 600\; \rm km/s$.

\nskip
When CRs couple to the plasma via scattering by MHD waves, the Galactic winds develop and CRs are picked up at the height $|z| \sim D/ V_{\rm w} $ by the wind with velocity $V_{\rm w}$. They are then transported out of the Galaxy (i.e: CRs will generally  not return). 
Since our simulations propagate test particles in a prescribed magnetic field and/or flow, we can {\em not} simulate the 
self-consistent launch of a CR-driven wind. Instead we use a simple analytical model.
\nskip
The velocity of a steady and axi-symmetric Galactic wind in the MHD approximation must take the form (e.g \citet{WeDa}):
\be
\label{vwind}
 \bm{V}_{\rm w} =   V_{\rm p} \: \frac{\bm{B}_{\rm p}}{|\bm{B}_{\rm p}|} + V_{\phi} \: \bm{\hat{\phi}}
\ee
Here $\bm{B}_{\rm p}$ is the poloidal magnetic field: $\bm{B}_{\rm p} = (B_{r} \: , \: 0 \: , \: B_{z})$. In our model we will neglect the motion in the
azimuthal ($\phi-$)direction since our model (including the CR source distribution) is axially symmetric, retaining only the wind component $V_{\rm p}$ along the poloidal field.
We do not employ a full model for the Galactic wind. Rather we assume that the poloidal wind speed varies with height $z$ above the disk as:
\be
\label{windspeed}
	V_{\rm p}(z) = V_{0} \: \left(\frac{|z|}{H_{\rm w}} \right) \; ,
\ee
a reasonable approximation sufficiently close to the disk for a wind accelerating away from the Galactic Disk. We use $H_{\rm w} = 20 \; {\rm kpc}$ in these simulations, and
vary $V_{\rm 0}$ between $0$ and $600$ km/s. The importance of CR advection by this wind is
determined by the dimensionless parameter:
\be
\label{windpar}
	\Xi_{\rm w} = \frac{V_{0} \: H_{\rm w}}{D_{zz}}  \simeq 10 \: \left( \frac{V_{0}}{100 \; {\rm km/s}} \right) \left( \frac{H_{\rm w}}{10 \; {\rm kpc}} \right)
	\left( \frac{D_{zz}}{3 \times 10^{28} \: {\rm cm^2/s}} \right)^{-1}\; .
\ee
Here $D_{zz}$ is the $zz$-component of the CR diffusion tensor. Advection away from the disk becomes the dominant transport 
mechanism for CRs when $\Xi_{\rm w} \gg 1$.
Of course, in this model (with $V_{\rm p} \propto |z|$) it is essential that diffusion first transports the CRs some distance away from the disk mid-plane.
As an illustration: if the CR is \lq{}picked up\rq{} by the wind at some height $h_{\ast} \ll H_{\rm w}$ from the mid-plane, the ratio of the diffusion time $t_{\rm diff} = H_{\rm w}^2/2 D_{zz}$
to a height $H_{\rm w}$ and the advection time $t_{\rm w} = (H_{\rm w}/V_{0}) \: \ln(H_{\rm w}/h_{\ast})$ to the same height is:
\be
	\frac{t_{\rm diff}}{t_{\rm w}} = \left( \frac{H_{\rm w} \: V_{0}}{2 D_{zz}} \right) \:
	\left[ \:  \ln \left(\frac{H_{\rm w}}{h_{\ast}} \right) \: \right]^{-1} =
	\frac{\Xi_{\rm w}}{2} \: \left[ \:  \ln \left(\frac{H_{\rm w}}{h_{\ast}} \right) \: \right]^{-1} \; .
\ee
In practice $h_{\ast}$ will roughly equal the thickness of the stellar disk of the Galaxy, $h_{\ast} \simeq 0.2-0.4 \; {\rm kpc}$.

%%%%%%%%%%%%%%%%%%%%%%%%%%%%%%%%%%%%%%%%%%%%%%%%%%%%%%%%%%%%%%%%%%%%%%%%%%%%%%
\subsection{Effective drift speed in the It\^o formulation}
\label{sec:2.5}

The precise treatment of drift and diffusion needs some discussion. 
Without scattering, the drift velocity of a charge $q$ with momentum $\bm{p}$ and velocity $\bm{v}$
in a static magnetic field $\bm{B}(\bm{x})$ is a combination of gradient $B$ drift and curvature drift, which equals (see Appendix A):
\be
\label{Vdnosc}
\bm V_{\rm gc} = \frac{cpv}{3q}\; \bm \nabla \times  \bigg(\frac{ \bm B}{B^{2}}\bigg) \; ,
\ee
when averaged over an isotropic distribution of momenta so that $<p_\perp \: v_{\perp}>/2 = < p_{\parallel} \: v_{\parallel} > = pv/3$, where
the brackets are the average over momentum direction and the subscript $\perp$ ($\parallel$) 
refers tho the component perpendicular (parallel) to the magnetic field.
This is the (slow) drift of the {\em guiding center}, the average position of the charge when one averages over the 
rapid gyration around the magnetic field. 
These drifts are fully discussed in the classic paper of \citet{Northrop}. The well-know $\bm{E} \btimes \bm{B}$ drift is included
in the wind velocity since the MHD condition applies so that $\bm{E} = - (\bm{V}_{\rm w} \btimes \bm{B})/c$.

\nskip
The {\em full} diffusion tensor, in a simple collisional model with collision frequency $\nus$, 
takes the form  in component notation (e.g. \citet{Miyam}, Ch. 7.3):
\be
\label{fullD}
	\mathrm{D}_{ij} = \Dpar \: b_{i} b_{j} + \Dperp \: \left( \delta_{ij} - b_{i} b_{j} \right) - D_{\rm a} \: \epsilon_{ijk} \: b_{k} \; .
\ee 
Here $b_{i}$ is the $i-th$  component of the unit vector $\hatb$ of the ordered magnetic field, and $\epsilon_{ijk}$ is the totally anti-symmetric symbol in three dimensions.
The three fundamental diffusion coefficients appearing in this expression are:
\be
\label{Dcomponents}
	\Dpar = \frac{v^2}{3 \nu_{\rm s}} \; \; , \; \; \Dperp = \Dpar \: \left(\frac{\nus^2}{\nus^2 + \Omega^2} \right) \; \; , \; \; 
	D_{\rm a} = D_{\parallel} \: \left( \frac{\nus \: \Omega}{\nus^2 + \Omega^2} \right)  \; .
\ee
Here $\Omega = q B/\gamma m c$ is the gyration frequency of the charge with $\gamma = 1/\sqrt{1 - v^2/c^2}$ its Lorentz factor.
When this diffusion tensor is used in the diffusion equation the term involving $D_{\rm a}$ leads to an advection term (and not to a diffusion term because of the
anti-symmetry of this term in the indices $i$ and $j$), with  an effective guiding center drift velocity equal to:
\be
	\bm{V}_{\rm gc} = \frac{cpv}{3q} \grad \btimes \left\{  \frac{\Omega^2}{\nus^2 + \Omega^2} \: \bigg(\frac{\bm B}{B^{2}}\bigg) \right\} \; .
\ee
In our simple model we assume that $\Dperp/\Dpar \equiv \epsilon$ is a constant, where (\ref{Dcomponents}) then yields $\epsilon = \nus^2/(\nus^2 + \Omega^2)$.
Then, in order to be consistent, the guiding center drift velocity must be defined as:
\be
\label{Vd}
 \bm{V}_{\rm gc} = (1- \epsilon) \; \frac{cpv}{3q}\; \bm \nabla \times  \bigg(\frac{ \bm B}{B^{2}}\bigg) \; .
\ee
It reduces to the standard (collisionless) form when $\epsilon \ll 1$ ($\nus \ll \Omega$) and vanishes for $\epsilon = 1$, the case of isotropic diffusion.
This is physically correct.
\nskip
The diffusive random step $\Delta \bm{x}_{\rm diff}$ in (\ref{TPmotion}) only involves $\Dpar$ and $\Dperp$, the two diffusion
coefficients that determine the symmetric part of the diffusion tensor that can be written in dyadic notation as 
$\bm{\mathrm{D}}_{\rm symm} \equiv D_{\parallel} \: \hatb \hatb + D_{\perp} \: \left( \bm{\mathrm{I}} - \hatb \hatb \right)$.
If there are gradients in the field direction or in the coefficients $\Dpar$ and $\Dperp$ 
one must -in the It\^o formulation (\ref{TPmotion}) of the equations- include the gradient drift velocity equal to:
$\bm{V}_{\rm gr} = \grad \bdot \bm{\mathrm{D}}_{\rm symm}$.
The total drift velocity  $\bm{V}_{\rm dr} = \bm{V}_{\rm gr} + \bm{V}_{\rm gc}$ becomes, 
writing the diffusion tensor as the sum of the symmetric and the anti-symmetric part 
$\bm{\mathrm{D}} \equiv \bm{\mathrm{D}}_{\rm symm} + \bm{\mathrm{D}}_{\rm a}$:
\be
\label{drifttotal}
	\bm{V}_{\rm dr}  =   
	\grad \bdot \bm{\mathrm{D}}_{\rm symm}
	+ (1- \epsilon) \; \frac{cpv}{3q}\; \bm \nabla \times  \bigg(\frac{ \bm B}{B^{2}}\bigg) = 
	\grad \bdot \left( \bm{\mathrm{D}}_{\rm symm} + \bm{\mathrm{D}}_{\rm a} \right) \; . \nonumber
\ee
This is
the \lq{}standard form\rq{} found in the mathematical literature on the the It\^o formulation. In Appendix A we give explicit
analytical expressions for the drift velocity in our adopted magnetic field.

%%%%%%%%%%%%%%%%%%%%%%%%%%%%%%%%%%%%%%%%%%%%%%%%%%%%%%%%%%%%%%%%%%%%%%%%%%%%%%%%%%%%%%%%%%%%

\section{Results of the simulations}
\label{sec:3}

We present results from our simulations for two different values of the ratio: 
$\epsilon= \Dperp/\Dpar$: $\epsilon = 0.01$ (strongly anisotropic diffusion) and $\epsilon = 0.5$ (mildly anisotropic diffusion). 
The diffusion coefficients $D_{\perp}$ and $D_{\parallel}$ are kept constant for a given CR energy. 
%%%%%%%%%%%%%%%%%%%%%%%%%%%%%%%%%%%%%%%%%%%%%%%%%%%%%%%%%%%%%%%%%%%%%%%%%%%%%%%%%%%%%%%%%%%%
\subsection{The effect of the drift}
\label{sec:3.1}

\begin{figure}
	\centering
	\includegraphics [width=\columnwidth]{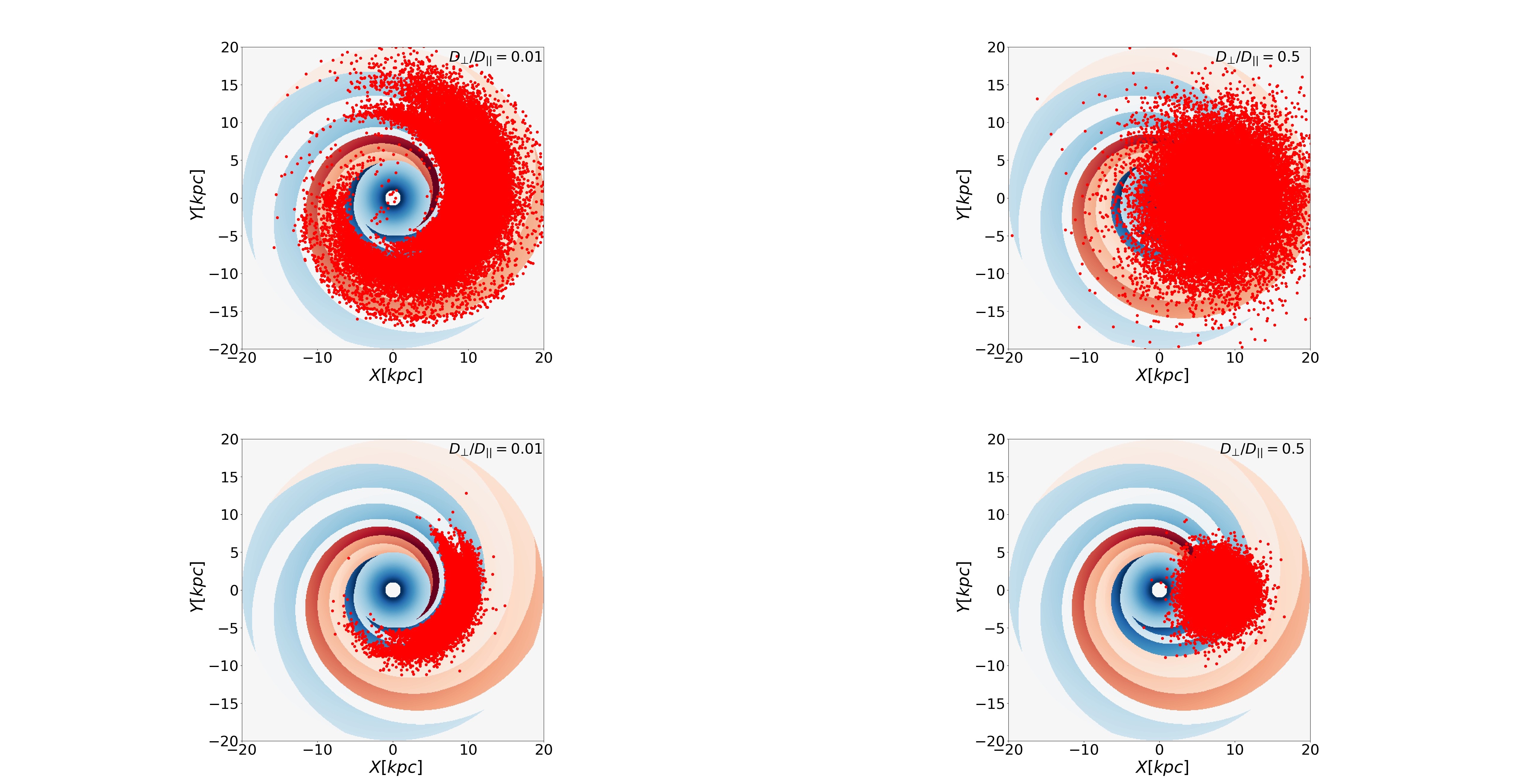}
	\caption{ The distribution  of CR protons projected onto the Galactic plane at the moment of escape. 
		The upper two plots show the results when the drift motion is not considered, in the lower two plots drift motion is taken into account.
		The right panels are for $\Dperp / \Dpar = 0.01$, the left panels are for $\Dperp / \Dpar = 0.5$.
		In these simulations we take $\Dpar = 3 \times 10^{28}$ cm$^{2}$ s$^{-1}$.
		The CRs were injected at $(X \: , \: Y) = (7 \; {\rm kpc} \: , \: 0)$.}
	%%%%%%%%%%%%%%%%%%%%%%%%%%%%%%%%%%%%%%%%%%%%%%%%%%%%%%%%%%%%%%%%%%%%%%%%%
	\label{figure1}
\end{figure}

Figure \ref{figure1} shows the  position  of CR protons, projected onto the Galactic plane, 
at the moment they reach the upper (lower) boundary of the CR halo, located at $z\;=+ H_{\rm cr}$ ($z=- H_{\rm cr} $) 
with $H_{\rm cr} = 4 \; {\rm kpc}$, or when they reach the outer radius of the Galaxy, taken to be $r_{\max} = 20 \; {\rm kpc}$.  
In these simulations there is no Galactic wind. All CRs were injected at $(X \: , \: Y) = (7 \; {\rm kpc} \: , \: 0)$.

For strongly anisotropic diffusion ($\Dperp/\Dpar=0.01$, the left two panels) the CRs are mostly follow the spiral field. 
In the mildly anisotropic case ($\Dperp/\Dpar=0.5$, the right two panels) CRs spread out almost isotropically from the injection site.
In the two top panels  the drift motion is neglected. In the bottom two panels the drift motion is taken into account.
Without drift, CR protons spread over a larger region of the disk before escaping. 
The drift motion leads to a faster escape of CRs, and as a result
compresses the distribution of the CRs. It also leads to a bulk inward drift to smaller radii. 
As the effective drift velocity is proportional to $\epsilon - 1 $, the effect of drift is smaller for
the case $\epsilon= 0.5$.

\begin{figure}
	\centering
	\includegraphics [width=\columnwidth]{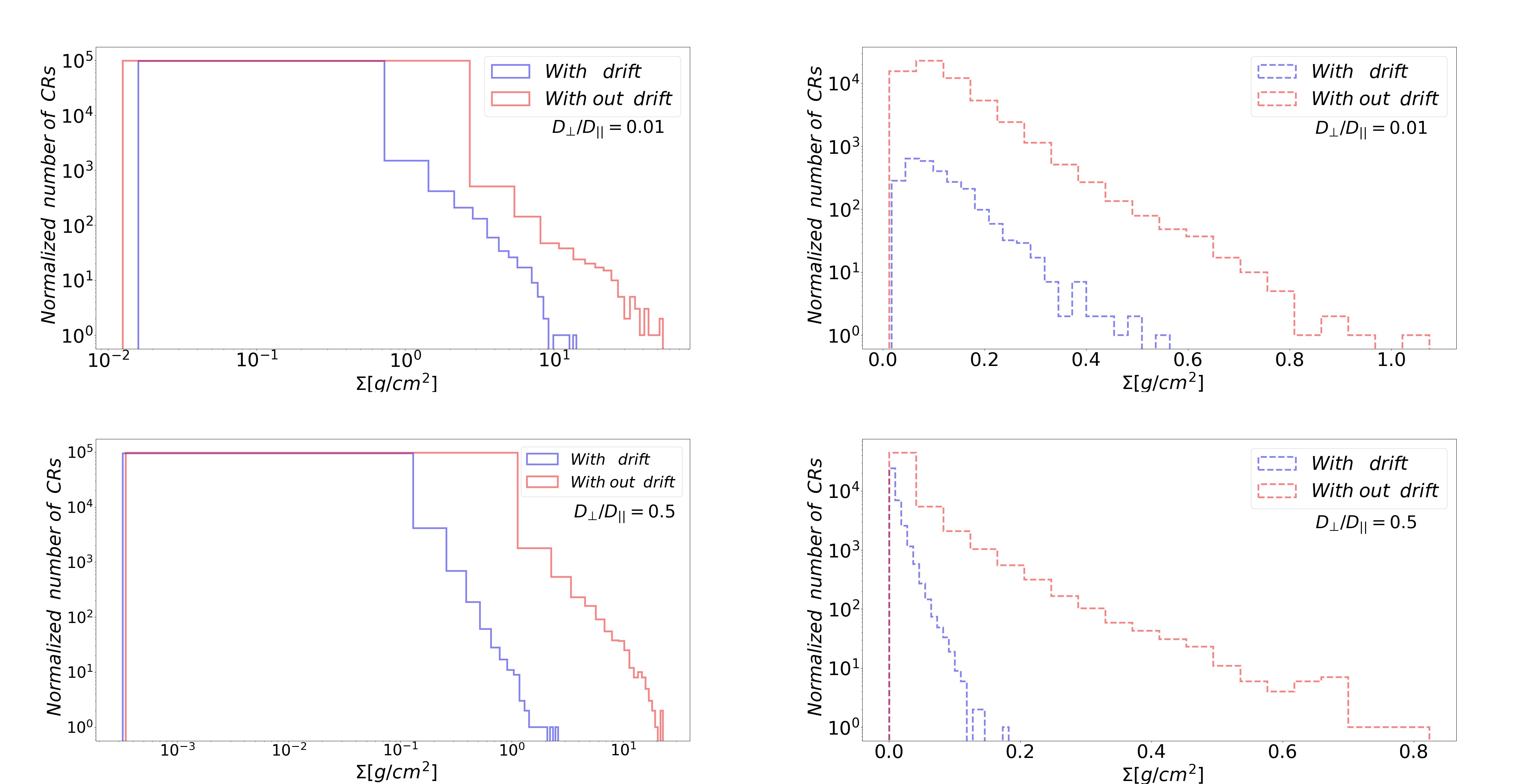}
	\caption{ The distribution of the accumulated  grammage. In the red histograms the drift motion is not considered in the calculation, 
		in the blue histograms  the drift motion is taken into account. The left column gives the grammage at the moment of escape, while the right
		shows it for CRs inside the local volume around the  Solar System. 
		Again $\Dperp/\Dpar=0.01$ and $\Dperp/\Dpar=0.5$, as indicated in each panel. As before $\Dpar = 3 \times 10^{28}$ cm$^{2}$ s$^{-1}$. }%
	\label{figure2}
\end{figure}

\nskip
Figure \ref{figure2} (left column) shows  the distribution of the CRs of  Figure \ref{figure1} over the accumulated grammage, 
calculated at the moment of escape from the Galaxy. 
In the red histogram the drift motion is  neglected, while in the blue histogram the drift motion is taken into account.
The right column of Figure \ref{figure2} shows the grammage distribution of these CRs observed around the Solar System,
without (in red) and with (in blue) drift. 
\nskip
Without drift, the accumulated grammage is larger as CRs spend more time in the CR halo. This allows them to 
spread out over a larger range in galactic radius before they escape. 
This agrees with the spatial distribution (projected onto the Galactic disk) shown in Figure \ref{figure1}. 
In conclusion:  given $\Dpar$ and $\epsilon$, the drift significantly decreases the residence time in the CR halo.

%%%%%%%%%%%%%%%%%%%%%%%%%%%%%%%%%%%%%%%%%%%%%%%%%%%%%%%%%%%%%%%%%%%%%%%%%%%%%%%%%
\subsection{CR advection by a Galactic wind}
\label{sec:3.2}

\begin{figure}
	\centering
	\includegraphics [width=0.8\columnwidth]{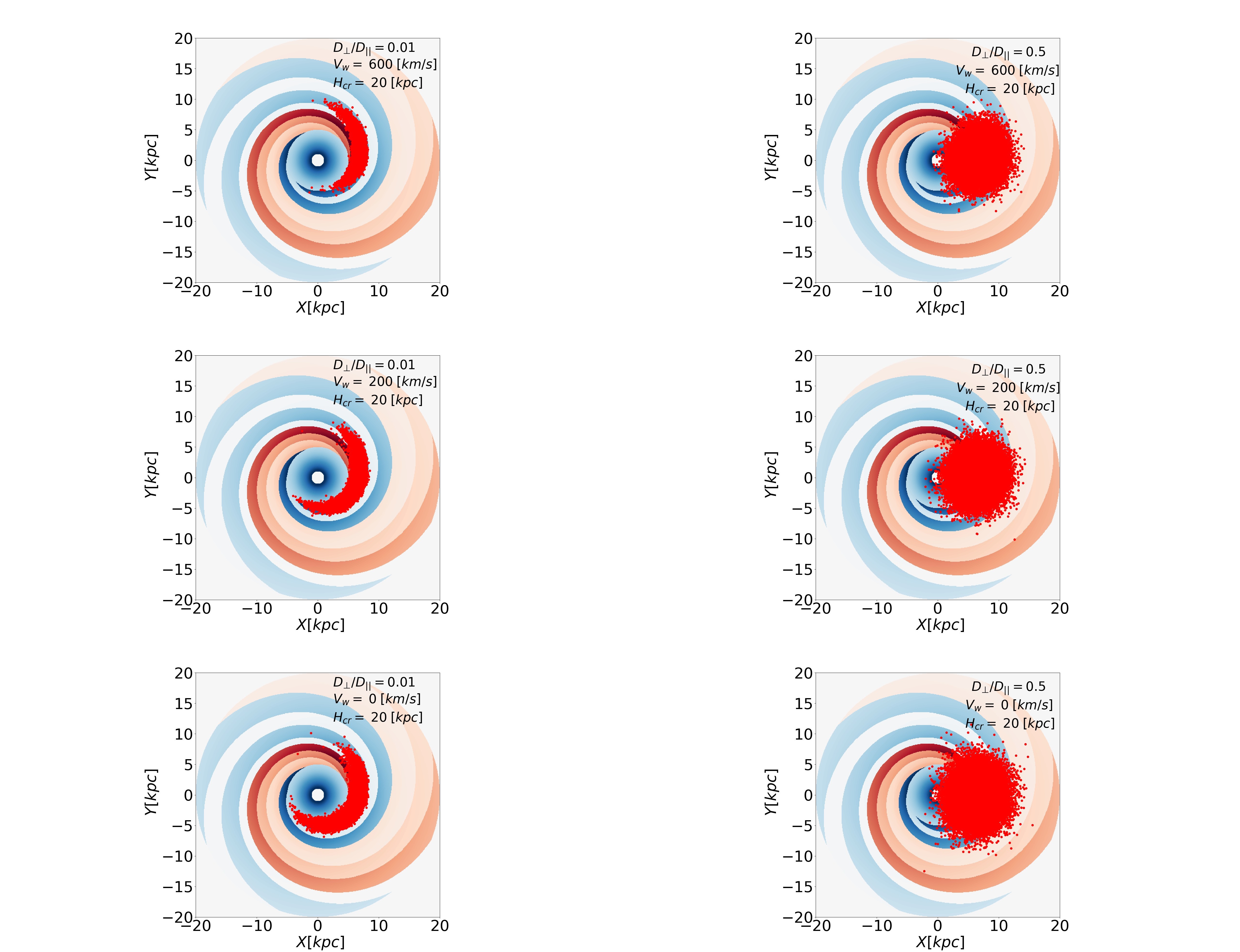}
	\caption{  The distribution of GCR protons projected onto the Galactic plane for $\Dperp/\Dpar=0.01$ (left) and $\Dperp/\Dpar=0.5$ (right). 
		The CRs were injected at $(X \: , \: Y) = (7 \; {\rm kpc} \: , \: 0)$.
		The drift motion and a Galactic wind are included in these results. 
		Results are for three different wind velocities, as indicated in each panel, 
		and take $\Dpar = 3 \times 10^{28}$ cm$^{2}$ s$^{-1}$. 
	}%
	\label{figure3}
\end{figure}

\begin{figure}
	\includegraphics[width=\columnwidth]{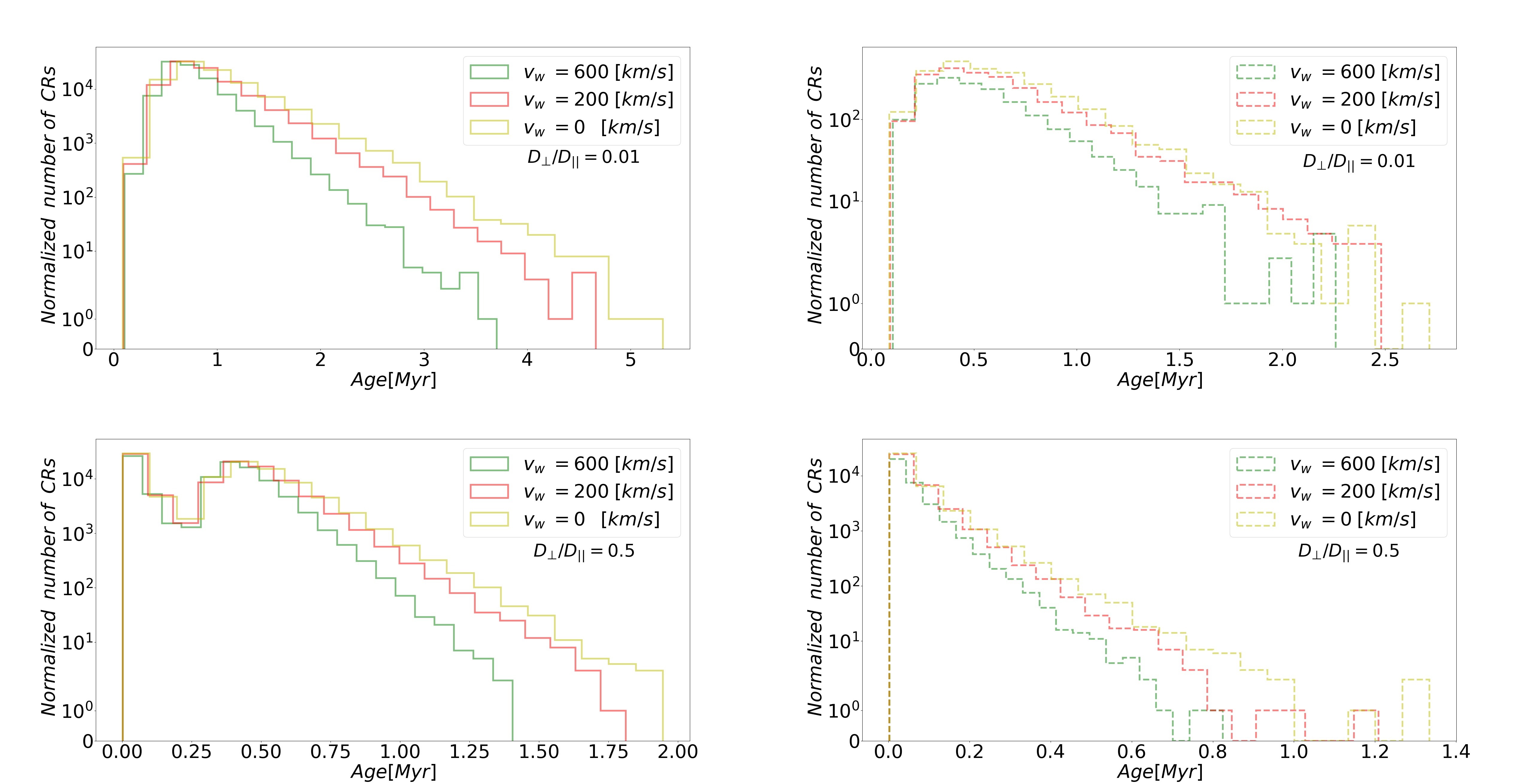}
	\caption{ The normalized  distribution over the age of CRs  at the moment of escape from the Galaxy (left column), while in the  right column is the normalized  distribution over the age of the CRs  inside the local volume around the Solar System. The drift motion and the Galactic wind are considered in the calculation. For three different wind velocities, and for  $\Dperp/\Dpar=0.01, 0.5$  as indicated in each panel. And take $\Dpar = 3 \times 10^{28}$ cm$^{2}$ s$^{-1}$.}%    
	\label{figure4}
\end{figure}

\begin{figure}
	\includegraphics[width=\columnwidth]{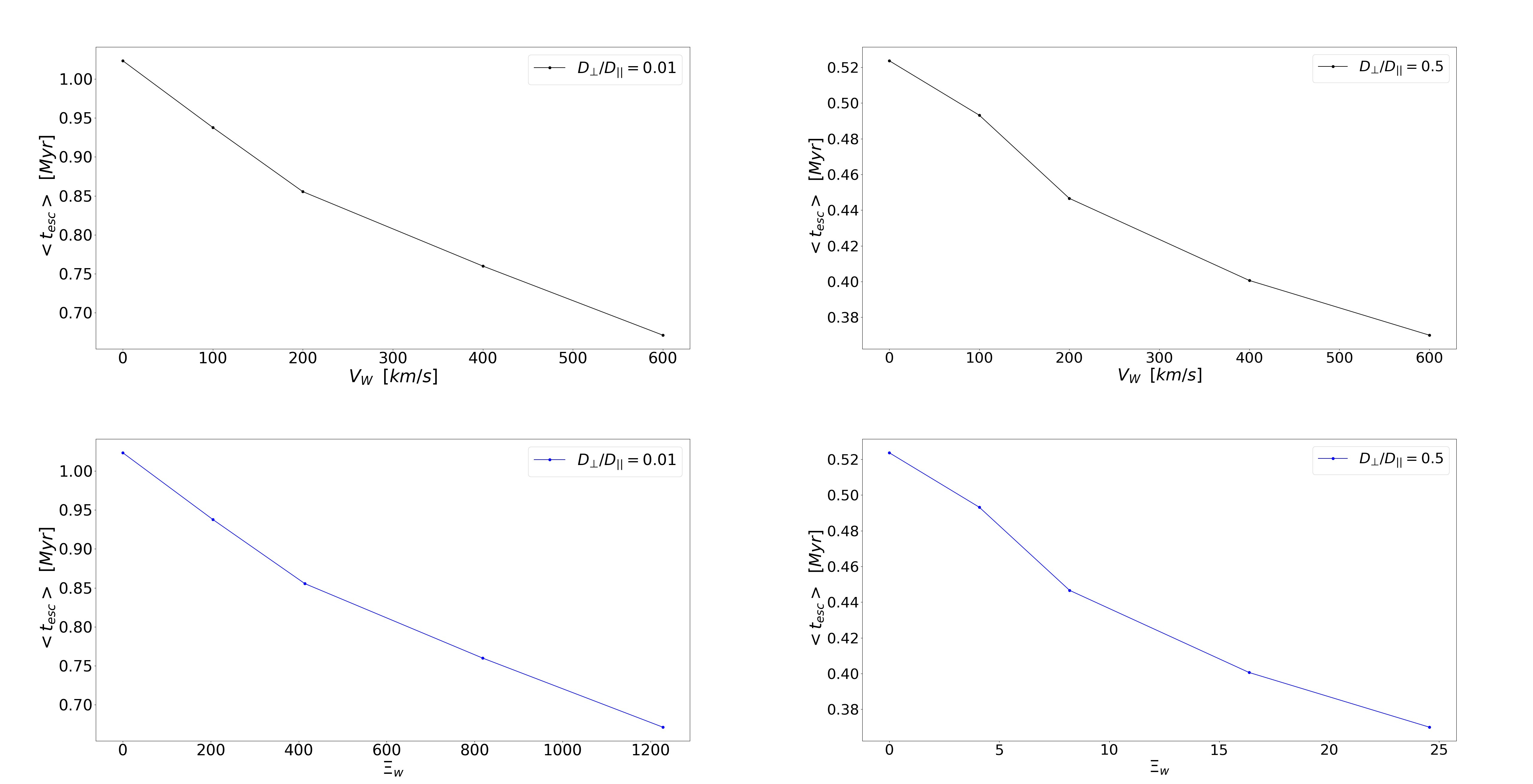}
	\caption{ The average CR age  at the moment of escape  as a function of wind velocity (fist row).  The second row shows the average CR age as a function of $\Xi_{\rm w}$ . For five different wind velocities, and for  $\Dperp/\Dpar=0.01, 0.5$ as indicated in each panel. In these simulations we take $\Dpar = 3 \times 10^{28}$ cm$^{2}$ s$^{-1}$. }%    
	\label{figure5}
\end{figure}

Figures 3 through 5 show the effect of CR advection by a Galactic wind.
The $\Dpar$  in these simulations is kept constant at $\Dpar = 3 \times 10^{28}$ cm$^{2}$ s$^{-1}$.
The wind velocity is taken to increase linearly with height $|z|$ away from the disk mid-plane, see prescription (\ref{windspeed}).
The resulting CR transport is diffusive close to the Galactic disk. It becomes convective further out, i.e. 
there is a strongly diminished chance that CRs return to the mid-plane of the Galactic disk.
We then extend the CR halo to a height $H_{\rm cr} = H_{\rm w} = 20 \; {\rm kpc}$ in these simulations.
\nskip
Figure \ref{figure3} (left column) shows the  position of  1 GeV CR protons projected onto the Galactic plane at the moment
of escape. The left row is for $\Dperp/\Dpar = 0.01$ and the right row for $\Dperp/\Dpar = 0.5$. 
The CRs where injected at $(X \: , \: Y) = (7 \; {\rm kpc} \: , \: 0)$.
We use three different velocities
(see Eqn. \ref{windspeed}): $V_{0} = 600\; \rm km/s$ (upper plot), $200\; \rm km/s$ (middle plot), and $0 \; \rm km/s$ (lower plot). In all cases one sees  outward CR transport, along the X-field lines perpendicular to the disk. 
Comparing the top two plots with the wind-less bottom plot it is evident that, with a wind, CRs escape sooner and -as a consequence- fan out less in both the radial and azimuthal directions. While  in the  right column the effect of the wind is much less evident.
\nskip
Figure \ref{figure4}, left column, shows  the normalized distribution in the age (residence time)
of the CR protons  at the moment they escape the Galaxy.
The right column shows the normalized distribution over the age of the CR protons that observed in the local volume of
$1 \ {\rm kpc}$ radius around the Solar System.
\nskip 
In Figure \ref{figure5}  we show the average CR age at the moment of escape  as a function of wind velocity (first row).  
The second row we show the average CR age as a function of $\Xi_{\rm w}$ as defined in Eqn. (\ref{windpar}). Now CRs are injected over the entire
Galactic disk. The CRs are given a weight $\propto N_{\rm snr}(r_{\rm inj})$, with $r_{\rm inj}$ the injection radius 
and $N(r)$ is the Galactic surface density of supernova remnants, taken to be the sources of these CRs.
We employ the SNR surface density given by \citet{Case1996}: 
\be
\label{W2}
	N_{\rm snr}(r) \propto  \left(\frac{r}{{R}_{\odot}} \right)^{\alpha} \;   {\rm exp}\left( - \frac{r}{R_{\rm snr}} \right) \; ,
\ee
where  ${R}_{\odot} =8.5 \; {\rm kpc}$  is the position of the Sun, $\alpha = 1.1$, and  $R_{\rm snr} \;= 8.0  \;{\rm kpc}$.
\nskip
We employ five values for the characteristic wind speed $V_{0}$:  between $0\; \rm km/s$ to $600\; \rm km/s$. 
In these simulations we take $\Dpar = 3 \times 10^{28}$ cm$^{2}$ s$^{-1}$,  and we
choose $\Dperp/\Dpar = 0.01$ and $\Dperp/\Dpar = 0.5$.

As clearly seen the  average age of CRs decreases as the wind velocity increases. Even though the escape boundary is now at
$H_{\rm cr} = 20 \: {\rm kpc}$, the typical residence time (shown in figure \ref{figure5}) is still around 1 Myr, comparable to what we find
in the simulations without a wind where we put $H_{\rm cr} = 4 \: {\rm kpc}$. In the pure diffusion case one would expect
an increase of the residence time ($\propto H_{\rm cr}^2$) by a factor $\sim 16$. This shows that advection by the wind rapidly
becomes important. 
The same behavior is seen if one plots
the average CR age as a function of $\Xi_{\rm w}$. In conclusion: given $\Dpar$, increasing the wind velocity leads to a 
reduction of the CR residence time in the Galaxy. 

\begin{figure}
	\centering
	\includegraphics [width=\columnwidth]{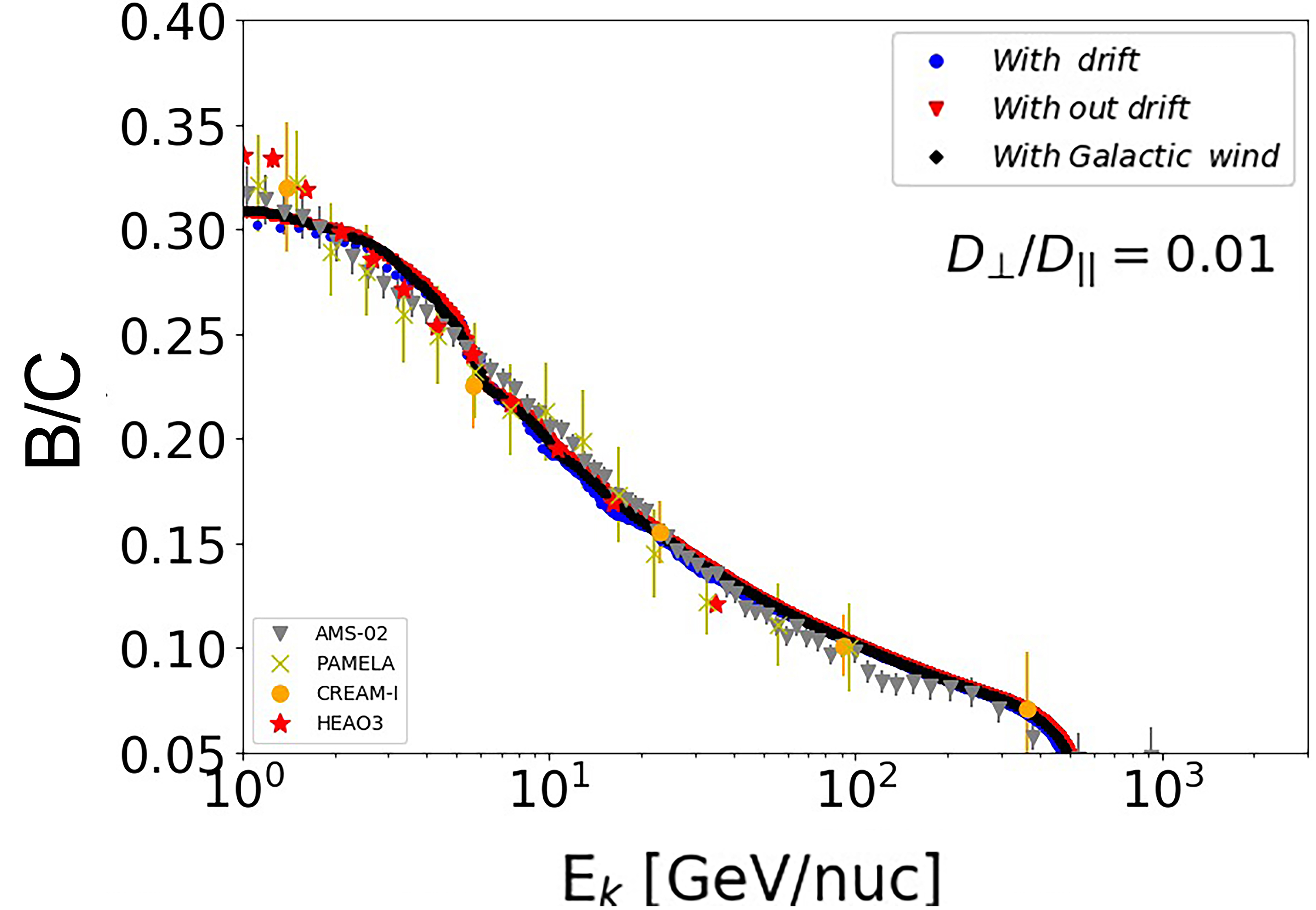}
	\caption{
		The ratio of  Boron over Carbon  abundance ratio  as a function of kinetic energy per nucleon.
		The gray triangles, yellow crosses, orange points, and red stars  are the results of measurements by  AMS-02 (\citet{AMS02-2016}), PAMELA (\citet{PAMELA2008}), CREAM (\citet{CREAM2008}), and  HEAO3 (\citet{HEAO31990}) respectively.
		The curves illustrate our results for the case of the diffusion coefficient $\Dperp/\Dpar=0.01$ once when the drift motion is  considered in the calculation (blue curve), the drift motion is not considered in the calculation(red curve), and the wind velocity is considered in the calculation (black curve). We use $D_{0} = 6.1 \times 10^{28}$ cm$^{2}$ s$^{-1}$ for the drift motion (blue curve),  with $D_{0} = 3 \times 10^{28}$ cm$^{2}$ s$^{-1}$ when the drift motion is neglected (red curve), and  $D_{0} = 2.7 \times 10^{29}$ cm$^{2}$ s$^{-1}$  for the Galactic wind ( black curve) to produce the best fit with the observational data.}  %
	
	\label{figure6}
\end{figure}

\nskip
Finally, in Figure \ref{figure6} we show the B/C ratio as a function of kinetic energy per nucleon. 
We used the weighted slab technique using the Path Length Distributions (PLDs) as described in \citet{ALZetoun2020}.
for  $\Dperp/\Dpar=0.01$.
In this Figure, the red curve is without drift motion, the blue curve takes drift motion into account,  while the black curve takes the wind velocity into account. The experimental data from  AMS-02 \citep{AMS02-2016}, PAMELA \citep{PAMELA2008}, CREAM \citep{CREAM2008}, and HEAO3 \citep{HEAO31990}  are shown for comparison.
It is possible to get a satisfactory agreement between  our results and observational data.
The parallel diffusion coefficient is assumed to scale with CR energy as $\Dpar = D_{0} \: \left({\cal R}/1 \: {\rm GeV}/c \right)^{\delta}$ with 
$\delta = 0.33$.
For the calculation without  drift (red  curve)  we  use  $D_{0}= 3 \times 10^{28}$ cm$^{2}$ s$^{-1}$, for the calculation with  drift (blue curve)  we  chose the value of $D_{0} = 6.1 \times 10^{28}$ cm$^{2}$ s$^{-1}$, and for the calculation with a wind we chose the value of  $D_{0} = 2.7 \times 10^{29}$ cm$^{2}$ s$^{-1}$ in order to  match the observed  B/C ratio at  $1\; \rm GeV/nucleon$.

%%%%%%%%%%%%%%%%%%%%%%%%%%%%%%%%%%%%%%%%%%%%%%%%%%%%%%%%%%%%
%%%%%%%%%%%%%%%%%%%%%%%%%%%%%%%%%%%%%%%%%%%%%%%%%%%%%%%
\section{ Conclusions }
\label{sec:4} 

In this paper we have investigated by means of numerical simulations the effect of drift motion, as well as the effect of 
advection of CRs away from the disc by a Galactic
wind on the propagation CRs in the Galaxy. We modified the magnetic field model of Jansson and Farrar, while retaining
essential features of this model, such as the spiral structure close to the mid-plane of the Galactic disk.
The main results are as follows:
\begin{itemize}
\item We show  that the drift  motion alone affects the transport of CRs  in the Galaxy, 
by compressing the CR distribution and by shifting them inward to smaller Galacto-centric radii;
\item We show  how a Galactic wind affects the transport of CRs  in the Galaxy, by advecting them away from their sources. This significantly
reduces (for given $\Dpar$ and $\epsilon = \Dperp/\Dpar$) the residence time in the Galaxy and the 
accumulated grammage, as expected from simple arguments.
This implies that, given the observed grammage derived from observations of (for instance) the B/C ratio, the diffusion coefficient
$\Dpar$ must increase for larger values of the Galactic wind speed in order to reproduce the observations. This implies that the
sources contributing to the CR flux at Earth must be closer compared to the case without a Galactic wind.
\item Away from the disk the flaring vertical X-field leads to a more rapid (mostly advective) transport of CRs to larger 
Galacto-centric radii when a wind is present.
\item As is the case without drift and wind, the accumulated grammage and the residence time depend strongly on the 
diffusion  ratio  $\Dperp/\Dpar$,  as already found in \citet{ALZetoun2018} for the case without drift or a wind.   
\end{itemize}

%%%%%%%%%%%%%%%%%%%%%%%%%%%%%%%%%%%%%%%%%%%%%%%
\section{ Data availability }
\label{sec:5}

The datasets generated during and/or analysed during the current study are available from the corresponding author on reasonable request.

\bibliographystyle{mnras}
\bibliography{Al_Zetoun_references}

\iffalse
\fi
%%%%%%%%%%%%%%%%%%%%%%%%%%%%%%%%%%%%%
\clearpage

%%%%%%%%%%%%%%%%%%%%%%%%%%%%%%%%%%%%%%%%%%%%%%%%%%%%%%%%%%%%%%%%%%%%%%%%%

%%%%%%%%%%%%%%%%%%%%%%%%%%%%%%%%%%%%%%%%%%%%%%%%%%%%%%End the GMF%%%%%%%%%%%%%%

%%%%%%%%%%%%%%%%%%%%%%%%%%%%%%%%%%%%%%%%%%%%%%%%%%%

%%%%%%%%%%%%%%%%%%%%%%%%%%%%%%%%%%%

%%%%%%%%%%%%%%%%%%%%%%%%%%%%%%%%%%%%%%%%%%%%%%%%%%%%%%%%%%%%%%%%%

%%%%%%%%%%%%%%%%%%%%%%%%%%%%%%%%%%%%%%%%%%%%%%%%%%%%%%%%%%%%%%%%%%%%%%%%%%%%%%%%%%%%%%%%%%%%%%%%%%%%%%%%
\clearpage
\appendix

\section{Guiding center and gradient drift velocities}

We briefly give the analytical results for the drift speeds as they apply in the simplified Jansson-Farrar field employed in this paper.

\subsection{Guiding center drift without scattering}

The motion of charged particles with charge $q$ in a non-uniform magnetic field $\bm{B} (\bm{x})$ and a sufficiently weak electric field $\bm{E}(\bm{x})$ (with $|\bm{E}| \ll |\bm{B}|$)
can be described as a combination of rapid gyration, fast motion along the magnetic field with velocity $v_{\parallel}$ and a (slow) 
drift motion of the guiding center (center of the gyro-orbit). The fast motion along the field
is subject to scattering and is taken into account by 
the parallel diffusion term with diffusion coefficient $\Dpar$. Here we concentrate on the slow drift.
% see \citet{Northrop}.

If we denote the position of the guiding center by $\bm{R}$, the drift velocity (to leading order) without scattering equals 
\be
\label{GCD}
\left(\frac{{\rm d} \bm{R}}{{\rm d}t} \right)_{\rm drift} \equiv \bm{V}_{\rm gc} = c \: \frac{\bm{E} \btimes \bm{B}}{B^2} + 
\frac{cpv}{3q} \left\{ \grad \btimes \left( \frac{\bm{B}}{B^2} \right) \right\} \; .
\ee 
We assume that there are no other (non-electromagnetic) forces acting on the charge, neglect the polarization drift, which is allowed for slow variations in the electric field. We also take the CR momentum distribution to be isotropic in momentum space. 
The first term is the well-known $\bm{E} \btimes \bm{B}$ drift. The second term is a combination of the drift due to the gradient
of the magnetic field strength, the curvature drift and the parallel drift. As such it is the average over solid angle in momentum space of
(see \citet{Northrop} for details)
\be
\frac{c p_{\perp} v_{\perp}}{2 qB^2} \: \left( \hatb \btimes \grad B \right) + 
\frac{c p_{\parallel} v_{\parallel}}{qB} \left( \hatb \btimes (\hatb \bdot \grad)\hatb \right) +
\frac{c p_{\perp} v_{\perp}}{2 q B} \: \left( \hatb \bdot (\grad \btimes \hatb) \right) \: \hatb \; .
\ee
with $p_\perp$ and $p_{\parallel}$ ($v_\perp$ and $v_{\parallel}$) respectively 
the components of momentum (velocity) perpendicular to and along the magnetic field. 
Then -on average- $<p_{\parallel} v_{\parallel}> = <p_{\perp} v_{\perp}>/2 = pv/3$ and one finds the second term in Eqn. (\ref{GCD}).
The $\bm{E} \btimes \bm{B}$ drift is included automatically if one allows for a bulk flow (wind) with velocity $|\bm{V}_{\rm w}| \ll c$ 
and uses the ideal MHD condition, $\bm{E} = - (\bm{V}_{\rm w} \btimes \bm{B})/c$. 
In that case has to interpret the particle momentum and velocity as those in the local rest frame of the bulk flow, 
and add the wind speed $\bm{V}_{\rm w}$ to the (average) CR velocity. This is what we do here. We neglect the small drift that results from the
fact that this rest frame is -generally speaking- not an inertial frame.

%%%%%%%%%%%%%%%%%%%%%%%%%%%%%%%%%%%%%%%%%%%%%%%%%%%%%%%%%%%%%%%%%%%%%%%%%%%%%%%%%%
\subsection{Guiding center drift with scattering}

As argued in the main paper the guiding center drift involves a reduced effective drift velocity
\be
\bm{V}_{\rm gc} = (1 - \epsilon) \:  \frac{cpv}{3q} \left\{ \grad \btimes \left( \frac{\bm{B}}{B^2} \right) \right\} \; ,
\ee
when scattering is important, with $\epsilon = D_{\perp}/D_{\parallel}$. 
This velocity can be rewritten as
\be
\label{gcvelocity}
\bm{V}_{\rm gc} = (1 - \epsilon) \: \frac{cpv}{3qBr} \: \bm{\Delta}_{\rm gc} \; ,
\ee 
where the dimensionless vector $\bm{\Delta}_{\rm gc}$ equals
\be
\label{Deltagc}
\bm{\Delta}_{\rm gc}  =  Br \: \left\{ \grad \btimes \left( \frac{\bm{B}}{B^2} \right) \right\} \; .
\ee
If we define a typical gyroradius by $r_{\rm g} = pc/qB$, the factor in front of $\bm{\Delta}_{\rm gc}$, which determines the typical guiding center speed,
can be written as
\be
\label{gcmagn}
(1 - \epsilon) \: \frac{cpv}{3qBr} = (1 - \epsilon) \: \left[ \frac{v}{3} \: \left( \frac{r_{\rm g}}{r} \right) \right] \; .
\ee

%%%%%%%%%%%%%%%%%%%%%%%%%%%%%%%%%%%%%%%%%%%%%%%%%%%%%%%%%%%%%%%%%%%%%%%%%%%%%
\subsection{Gradient drift}

For constant $D_\parallel$ and $D_\perp$ there is a gradient drift due to changes in the direction of the magnetic field.
The associated velocity $\bm{V}_{\rm gr} = \grad \bdot \bm{\mathrm{D}}_{\rm symm}$ is
\be
\label{graddriftspeed}
\bm{V}_{\rm gr} = D_{\parallel} (1 - \epsilon) \: \grad \bdot \left(\hatb \: \hatb \right) \; .
\ee
It can be written as
\be
\label{grd}
\bm{V}_{\rm gr} = (1 - \epsilon) \:\left[ \frac{v}{3} \left( \frac{\lambda_{\rm s}}{r} \right) \right] \: \bm{\Theta} \; ,
\ee
with the dimensionless vector $\bm{\Theta}$ defined as
\be
\label{thetadefroot}
\bm{\Theta} = r \: \grad \bdot \left(\hatb \: \hatb \right) = r \: (\bm{B} \bdot \grad) \left(\frac{\bm{B}}{B^2} \right) \; .
\ee
Here we used $D_{\parallel} = \lambda_{\rm s} v/3$ with $\lambda_{\rm s} = v/\nu_{\rm s}$ the parallel scattering length, 
employed $\hatb = \bm{B}/B$ and $\grad \bdot \bm{B} = 0$.
The factor in front of $\bm{\Theta}$ in relation (\ref{grd}) gives the typical magnitude of the gradient drift speed. 
Comparing this with the guiding center drift
speed (\ref{gcmagn}) one finds that
\be
\frac{|\bm{V}_{\rm gr}|}{|\bm{V}_{\rm gc}|} \simeq \frac{\lambda_{\rm s}}{r_{\rm g}} \; .
\ee
The two speeds have a similar magnitude when the parallel scattering mean-free-path becomes comparable with the gyro radius, the case
of {\em Bohm diffusion} where CR diffusion is almost isotropic. Strongly anisotropic diffusion occurs when $\lambda_{\rm s} \gg r_{\rm g}$,
in which case $|\bm{V}_{\rm gr}| \gg |\bm{V}_{\rm gc}|$.
%%%%%%%%%%%%%%%%%%%%%%%%%%%%%%%%%%%%%%%%%%%%%%%%%%%%%%%%%%%%%%%%%%%%%%%%%%%%%%%%%%%
\subsection{Velocities in the simplified JF field}

Table A1 below give the parameters needed to calculate the guiding center drift and the gradient drift. It lists the components of
$\bm{\Delta}_{\rm gc} \equiv (\Delta_{r} \: , \: \Delta_{\phi} \: , \ \Delta_{\theta})$ and of 
$\bm{\Theta} \equiv (\Theta_{r} \: , \: \Theta_{\phi} \: , \: \Theta_{z})$. The table lists the results for $z \ge 0$. For $z < 0$ both
$\Delta_{z}$ and $\Theta_{z}$ have the opposite sign. In these expressions we use for $|z| \le h$ the parameters
$B_{\rm D} = B_{{\rm D} 0}(r/r_{0})$ and 
$B_{\rm X}(r) = (B_{0}^{\rm X} \: \sin i) \: {\rm exp}(-r_{\rm p}/H_{\rm X})$ and $B = \sqrt{B_{\rm D}^2 + B_{\rm X}^2}$. 
The value of $B_{{\rm D} 0}$ at $r_{0} = 5 \: {\rm kpc}$ is listed in Table 1 of \citet{JF2012A} for all the spiral sections of the disk field.
Also: $B{_{0}^{\rm X}} = 4.6 \: \mu {\rm G}$.
The inclination angle $i$ of the X-field and the radius $r_{\rm p}$ have been defined above. Again for $|z| \le h$ we define the quantity
\be
\label{der1}
  \Lambda(r \: , \: r_{\rm p})  \equiv - \frac{r}{B_{z}} \: \frac{{\rm d} B_{z}}{{\rm d}r} 
  = 
  \left\{ \begin{array}{ll} {\displaystyle  \frac{r_{\rm p}}{H_{\rm X}} + \cos^2 i } & 
  	\mbox{for $r_{\rm p} \le r_{\rm X}$,} \\
  	& \\
  	{\displaystyle  \frac{r}{H_{\rm X}} } & 
  	\mbox{for $r_{\rm p} > r_{\rm X}$.} \\
  \end{array} \right. 
\ee
The length parameters appearing in Table A1 are: $h = 0.4 \: {\rm kpc}$, $H_{\rm X}= 2.9 \: {\rm kpc}$ and $r_{\rm X} = 4.8 \: {\rm kpc}$, taken from \citet{JF2012A} and \citet{JF2012B}. 

\begin{table}
          %\color{red}
	\centering
	\begin{tabular}{|c|c|c|} \hline
		\multicolumn{3}{|c|}{{\bf Table A1}}\\
		\multicolumn{3}{|c|}{Parameters for the drift calculation: $|z| \le h = 0.4 \: {\rm kpc}$ }  \\ \hline \hline
		& $r_{\rm p} \le r_{\rm X}$ & $r_{\rm p} > r_{\rm X}$ \\ \hline
		& & \\
		$\Lambda(r \: , \: r_{\rm p})$ & ${\displaystyle \left( \frac{r_{\rm p}}{H_{\rm X}} + \cos^2 i \right)}$ & 
		$ {\displaystyle \frac{r}{H_{\rm X}}} $ \\
		& & \\
		$\Delta_{r}$ & $0$ & 0\\
		& & \\
		$\Delta_{\phi}$ & \multicolumn{2}{|c|}{
			${\displaystyle - \frac{ 2 B_{\rm X}}{B} \times \left\{ \: 
				\left( \frac{B_{\rm D}}{B} \right)^2 + \Lambda(r \: , \: r_{\rm p}) \: \left[ \left( \frac{B_{\rm X}}{B} \right)^2 - \frac{1}{2} \right]
				\right\}}$} 
		\\
		& & \\
		$\Delta_{z}$ & \multicolumn{2}{|c|}{ ${\displaystyle 
				\frac{ 2 B_{\rm D} \: \cos p}{B} \times \left\{ \: 
				\left( \frac{B_{\rm D}}{B} \right)^2 + \Lambda(r \: , \: r_{\rm p}) \: \left( \frac{B_{\rm X}}{B} \right)^2 
				\:  \right\}}$} \\
		& & \\
		$\Theta_{r} $& \multicolumn{2}{|c|}
		{  ${\displaystyle 2 \left( \frac{B_{\rm D}}{B} \right)^2 \left\{ \left[ \left(\frac{B_{\rm D} \: \sin p}{B} \right)^2 - \frac{1}{2}\right] 
				+ \Lambda(r \: , \: r_{\rm p}) \: \left(\frac{B_{\rm X}}{B} \right)^2 \: \sin^2 p  \right\}} $} \\
		& & \\
		$\Theta_{\phi} $& \multicolumn{2}{|c|}
		{  ${\displaystyle \frac{2 B_{\rm D}^2 \: \sin p \: \cos p}{B^2} \: \left\{
				\left( \frac{B_{\rm D}}{B} \right)^2 + \Lambda(r \: , \: r_{\rm p}) \: \left( \frac{B_{\rm X}}{B} \right)^2 
				\right\}} $} \\
		& & \\
		$\Theta_{z} $& \multicolumn{2}{|c|}
		{ ${\displaystyle \frac{2 B_{\rm D} \: B_{\rm X} \: \sin p}{B^2} \:  \times
				\left\{ \: 
				\left( \frac{B_{\rm D}}{B} \right)^2 + \Lambda(r \: , \: r_{\rm p}) \: \left[ \left( \frac{B_{\rm X}}{B} \right)^2 - \frac{1}{2} \right]
				\right\}}$} \\
		& & \\
		\hline \hline
		\multicolumn{3}{|c|}{Parameters for the drift calculation: $|z| > h = 0.4 \: {\rm kpc}$ }  \\ \hline
		& $r_{\rm p} \le r_{\rm X}$ & $r_{\rm p} > r_{\rm X}$ \\ \hline
		& & \\
		$\Delta_{r}$ & $0$ & 0\\
		& & \\
		$\Delta_{\phi}$ & ${\displaystyle- \frac{r_{\rm p} \: \sin i}{H_{\rm X}} \: \left(1 + \cos^2 i \right) } $ &
		${\displaystyle - \frac{r \: \sin i_{0}}{H_{\rm X}} \: \left(1 + \cos^2 i_{0} \right)}$ \\
		& & \\
		$\Delta_{z}$ & 0 & 0 \\
		& & \\
		$\bm{\Theta}$ & $2 \cos i \: \hatb$ & $ \cos i_{0} \: \hatb $ \\
		\hline
	\end{tabular}
\end{table}

%%%%%%%%%%%%%%%%%%%%%%%%%%%%%%%%%%%%%%%%%%%%%%%%%%%%%%
\end{document}